\documentclass[a4paper,11pt]{article}
\usepackage{pos}
\usepackage{graphicx}  % needed for figures
\usepackage{dcolumn}   % needed for some tables
\usepackage{bm}        % for math
\usepackage{amsmath}
\usepackage{epsfig}
\usepackage{rotating}
\usepackage{xspace}
\usepackage{hyperref}
\usepackage{xcolor}
\newcommand{\bea}{\begin{eqnarray}}
\newcommand{\eea}{\end{eqnarray}}

\title{Complex Langevin boundary terms in lattice models}
\ShortTitle{Complex Langevin boundary terms}

\author[a]{Michael Westh Hansen}
\author[c]{Erhard Seiler}
\author*[a]{D\'enes Sexty}
\author[b]{Ion-Olimpiu Stamatescu}

\affiliation[a]{Institute of Physics, NAWI Graz, University of Graz,\\
Universitätsplatz 5, Graz, Austria }

\affiliation[b]{Institut für Theoretische Physik, Universität Heidelberg,\\
  Philosophenweg 16, 69120 Heidelberg, Germany}

\affiliation[c]{Max-Planck-Institut für Physik (Werner-Heisenberg-Institut),\\
Föhringer Ring 6, München, Germany}

\emailAdd{michael.hansen@uni-graz.at}
\emailAdd{ehs@mpp.mpg.de}
\emailAdd{denes.sexty@uni-graz.at}
\emailAdd{I.O.Stamatescu@thphys.uni-heidelberg.de}

\abstract{In complex Langevin simulations, the insufficient decay of the probability density near infinity leads to boundary terms that spoil the formal argument for correctness. We present a formulation of this term that is cheaply measurable in lattice models, and in principle allows also the direct estimation of the systematic error of the CL method. Results for a toy model, 3d XY model and HDQCD are presented.}

\FullConference{%
 The 38th International Symposium on Lattice Field Theory, LATTICE2021
  26th-30th July, 2021
  Zoom/Gather@Massachusetts Institute of Technology
}

\begin{document}
\maketitle

\section{Introduction}

The notorius sign problem invalidates importance sampling simulations of
theories with a complex measure, e.g. QCD at nonzero chemical potential.
A proposed solution to this problem is the complex Langevin method
\cite{klauderparisi} which complexifies the integration manifold using holomorphy,
and uses the Complex Langevin equation (CLE), to evade the need of
interpreting the integration measure as a probability density.
For some complex measure $ \rho(x)=\exp(-S(x)) $ depending on the variable $x$ the CLE is thus written as 
\bea \label{CLE}
\partial_\tau x = - \textrm{Re} \left.{ \partial S(z) \over \partial z} \right|_{z=x+iy} +
\eta_\tau, \qquad
\partial_\tau y = - \textrm{Im} \left.{ \partial S(z) \over \partial z} \right|_{z=x+iy}, 
\eea
with the drift term $ K(z) = \partial S(z) / \partial z $ and a Gaussian noise $\eta$ satisfying $ \langle \eta_\tau \eta_{\tau'} \rangle=2 \delta(\tau-\tau') $.
This method is succesful in many cases \cite{success},
(for gauge theories there is an extra difficulty caused by
the complexification of gauge degrees of freedom, which is cured by
gauge cooling\cite{gaugecooling}). In some cases however problems remain,
leading to convergence to incorrect results.
It has been identified that the problematic cases either have to do
with insufficiently fast decay of the probability density of the
complexified stochastic process at infinity \cite{trust} or near zeroes of the measure \cite{pole}. In the formal justifiaction of the Complex Langevin method,
this gives rise to certain boundary terms invalidating the equivalence
of the Complex Langevin result to the correct result, which is simply
given by the integral on the original manifold with the complex measure.
Here we discuss boundary terms arising at infinity, for boundary terms
around poles, see \cite{seiler}.

\section{Boundary terms}

The CLE gives rise to a real probability density
on the complexified manifold
$P(x,y,\tau)$. (To keep the notation simple we use a one variable model,
generalizations to more variables are straightforward).
The CLE result for a holomorphic observable $O(z)$ is thus
\bea
  \langle O \rangle_{P(t)} = \int dx dy P(x,y,t) O(x+iy),
\eea
whereas the correct result we intend to calculate is
\bea
\langle O \rangle_\rho = \int dx O(x) \rho(x)=
\lim_{t\rightarrow \infty} \langle O \rangle _{\rho(t)} =
\lim_{t\rightarrow \infty} \int dx O(x) \rho(x,t)
\eea
where we defined a time-dependent complex measure $ \rho(x,t)$ using
$ \rho(x,t) = \exp(L_c^T) \rho_0(x)$, where $\rho_0(x)$ can be some initial
distribution on the real axis,
and $L_c$ is the complex Fokker-Planck operator $L_c= ( \partial_z + K(z)) \partial_z$.
Assuming the uniqueness of the limit $ \rho(x,t) \rightarrow e^{-S(x)} $,
one can define an interpolation function
\bea
F(t,\tau) = \int dx dy P(x,y,t-\tau) O(x,y,\tau),
\eea
using the time evolved observables
\bea
O(x,y,t) = e^{t L_c} O(x+iy).
\eea
$F(t,\tau)$ interpolates between the above two quantities
such that $F(t,0) =\langle O \rangle_{P(t) }$ and
$ F(t,t) = \langle O \rangle_{\rho(t)} $, provided the initial
conditions are chosen to agree. Thus the CLE results are
correct if $ \partial F(t,\tau) / \partial \tau=0$, which can be shown
using partial integrations assuming holomorphy of the evolved observables.
However the partial integrations can give rise to boundary terms
if the decay of $P(x,y)$ is not fast enough at infinity.
Introducing a cutoff in the integral we can derive the formula for the boundary
term at infinity:
\bea
B(Y,t) = \left. \partial_\tau F(Y;t,\tau) \right|_{\tau=0} =
\int_{|y|<Y} dx dy \partial_y ( K_y O(0) P(x,y,t) )  = \int_{y=Y} {\bf n} K_y P(t) O(0) dx dS
\eea
where the integral over a divergence is converted to an integral over a surface with $dS$ the surface element and $\bf n$ is a normal vector. Using this definition the boundary term for a toy model
was measured in \cite{scherzer1}. This definition allows the calculation
of the boundary term also in models with many independent variables, by defining a cutoff which restricts the variables to a compact subspace which encompasses the whole complexified manifold
as the cutoff $Y$ is sent to infinity. However the integration on the surface of such a compact region can be cumbersome for models with many variables. Instead we turn to different formulation of the boundary terms.

Following \cite{scherzer2}, starting again from $ B(Y,t)= \partial_\tau F(Y;t,\tau)  $, we write
\bea
\left. \partial_\tau F(Y;t,\tau)\right|_{\tau=0} =
- \int _{|y|\le Y} \partial_t P(x,y,t) O dx dy + \int_{|y| \le Y } P(x,y,t) L_c O dx dy.
\eea
The first term in this expression goes to zero in the $ t \rightarrow \infty $ limit, as the
process equilibrates and $P(x,y,t)$ evolves to a stationary solution. The second term
can be nonzero and it can spoil correctness if $ \lim _{Y\rightarrow \infty} B(Y) \neq 0 $.
We can similarly define higher order boundary terms using
\bea \label{Bncutoff}
B_n(Y) = \left. \partial_\tau^n F(Y;t,\tau)\right|_{\tau=0}=
\lim_{t\rightarrow \infty}  \int_{|y| \le Y } P(x,y,t) L_c^n O dx dy
= \langle \Theta ( Y -|y|) L_c^n O \rangle_P.
\eea
This construction can be straightforwardly generalized to lattice systems with many variables
by defining an integration cutoff with e.g. $ \max_i y_i < Y $. Generalization to
curved manifolds such as the SL$(3,\mathbb{C})$ space arising in the complexification
of lattice QCD simulations is e.g. possible using the unitarity norm
\bea
   n(M) = \textrm{Tr} (M^\dagger M -1 )^2 \textrm{ for } M \in \textrm{SL}(N,\mathbb{C}).
\eea
The boundary term is then defined using the Haar measure $dM$ as
\bea
 B_n(Y) = \int_{n(M)<Y} P(M) L_c^n O d M = \langle \Theta\left(Y-n(M)\right) L_c^n o \rangle_P.
\eea
For lattice models one can take $n(M)$ as the average or the maximum of the
univarity norms of the link variables.

The boundary term arises in the limit that the cutoff is taken to infinity, as well as the Lagevin
time. The order of limits is however crucial. As we see above $ B (Y=\infty)=0 $ expresses
the stationarity of the complexified process, and thus $ B (Y=\infty) $ is consistent with zero within errors
(large fluctuations give a hint for incorrect CLE results).
If we take the $Y\rightarrow \infty $ limit last, the observation of a nonzero boundary term
is possible.

\section{Results}

\begin{figure}[ht]
\begin{center}
  \includegraphics[width=0.48\columnwidth]{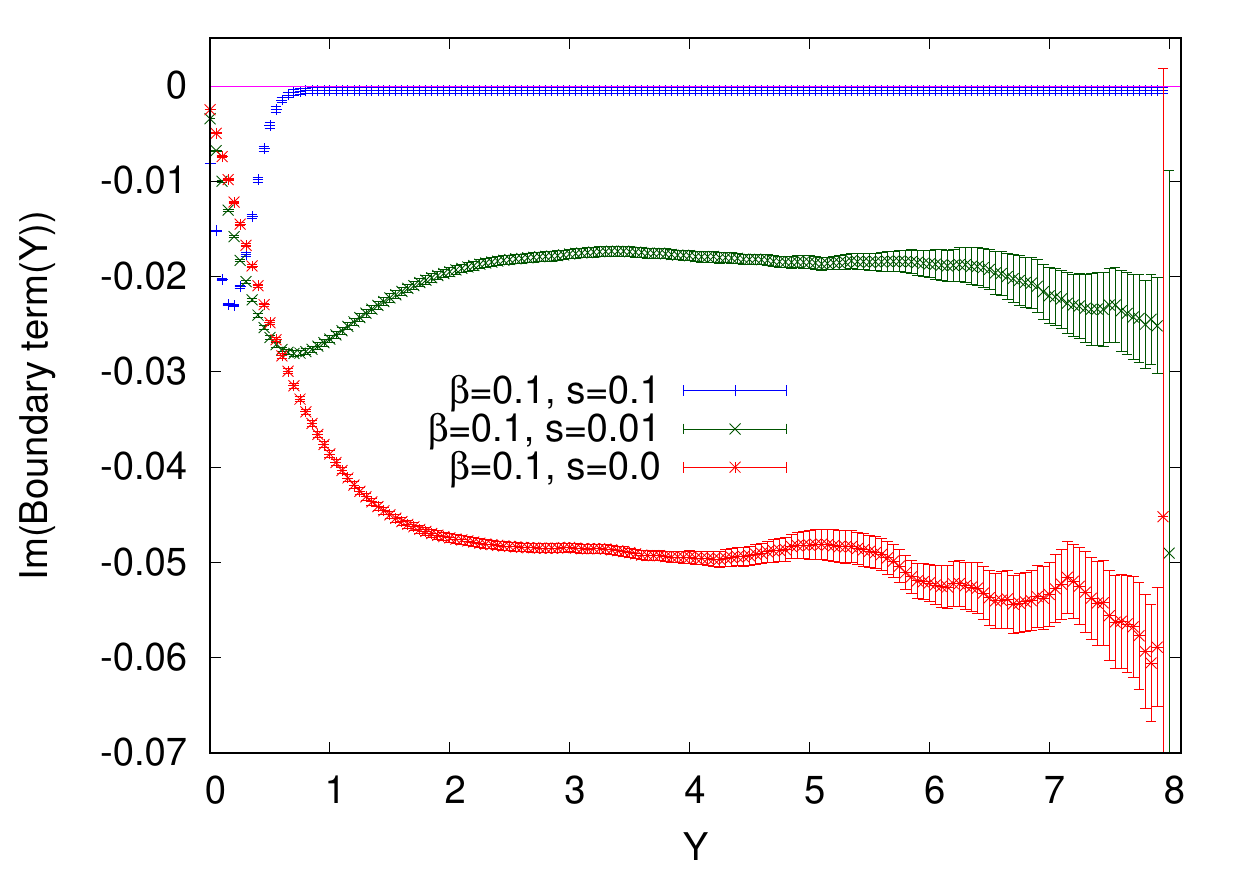}
  \includegraphics[width=0.48\columnwidth]{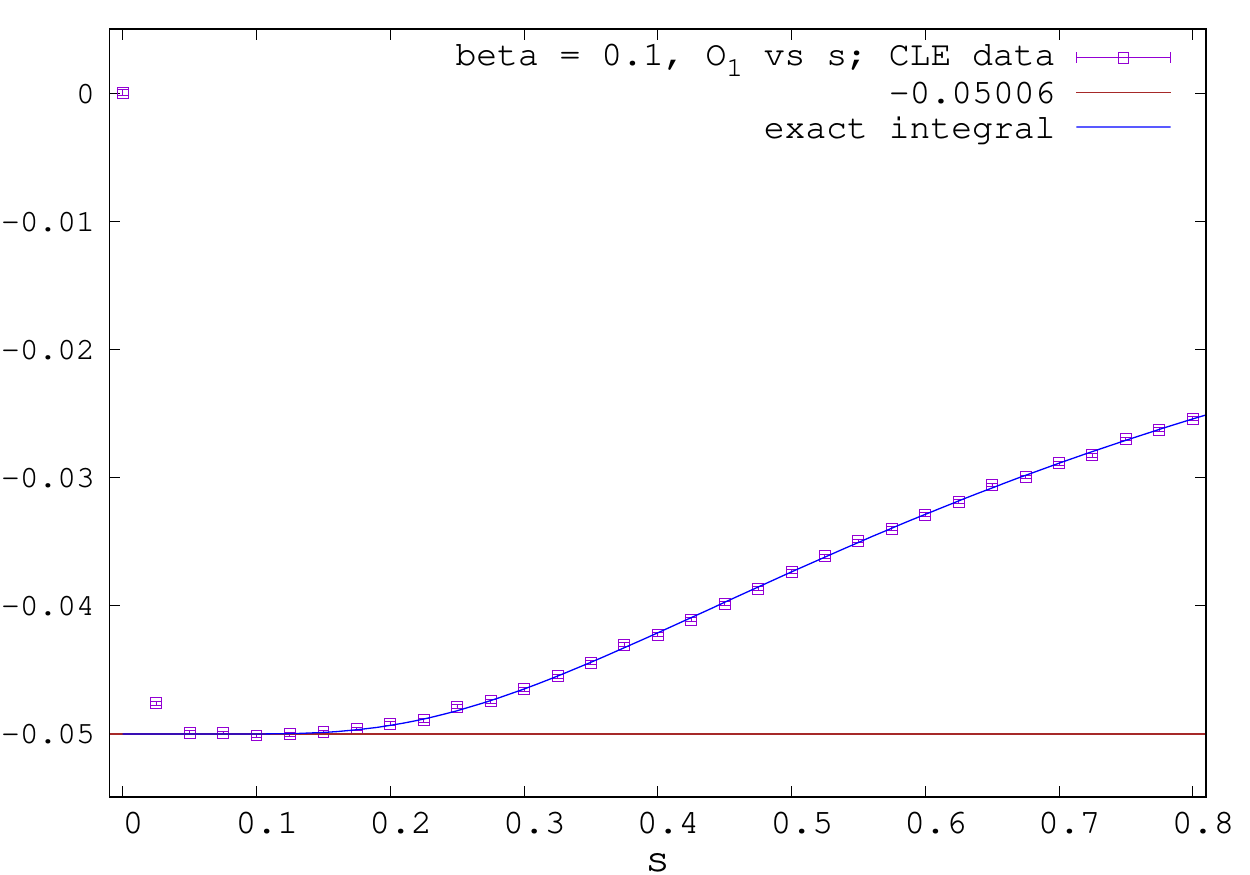}
%\vglue-20mm
\caption{Boundary term for the observable $e^{ix} $
  in the toy model $S(\varphi) = i \beta \cos(\varphi) + s \varphi^2/2$, for
  various $s$ values (left). The average $ \langle e^{ix} \rangle $ in the CLE process and its
  exact value as a function of $s$ for $\beta=0.1$ (right).
 }
\label{BT1}
\end{center}
\end{figure}

\begin{figure}[ht]
\begin{center}
  \includegraphics[width=0.48\columnwidth]{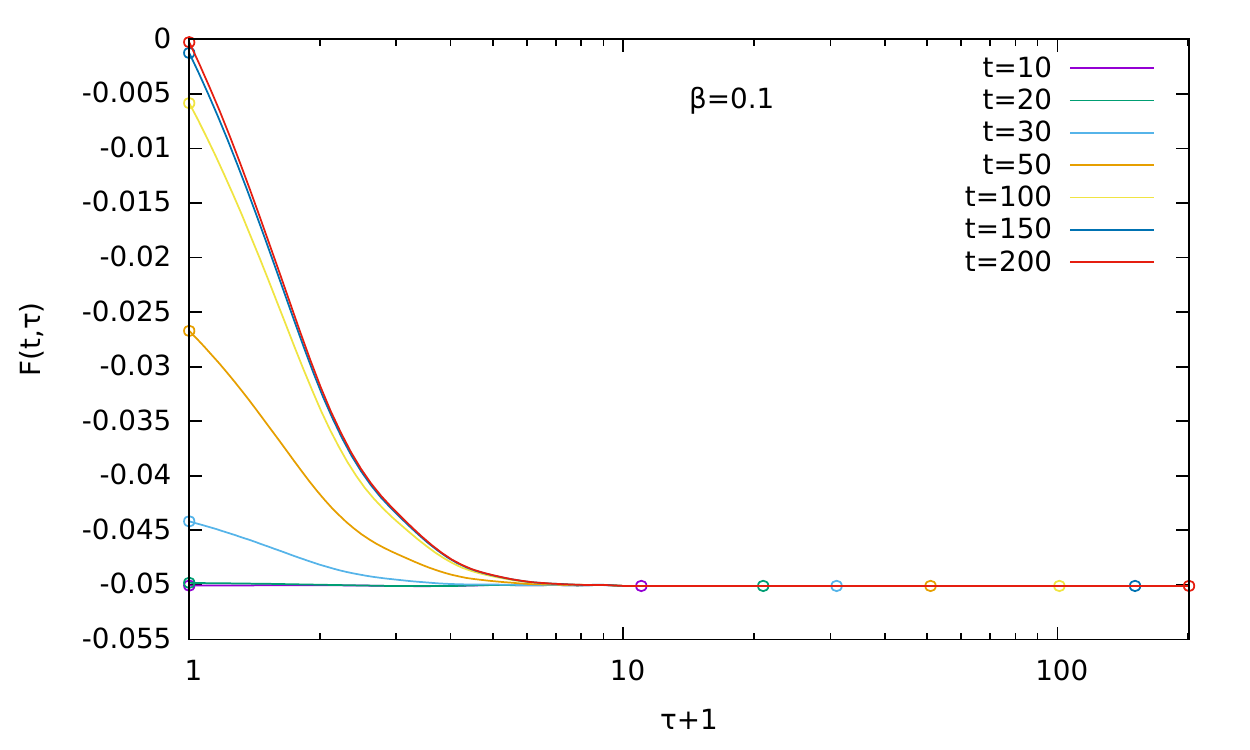}
  \includegraphics[width=0.48\columnwidth]{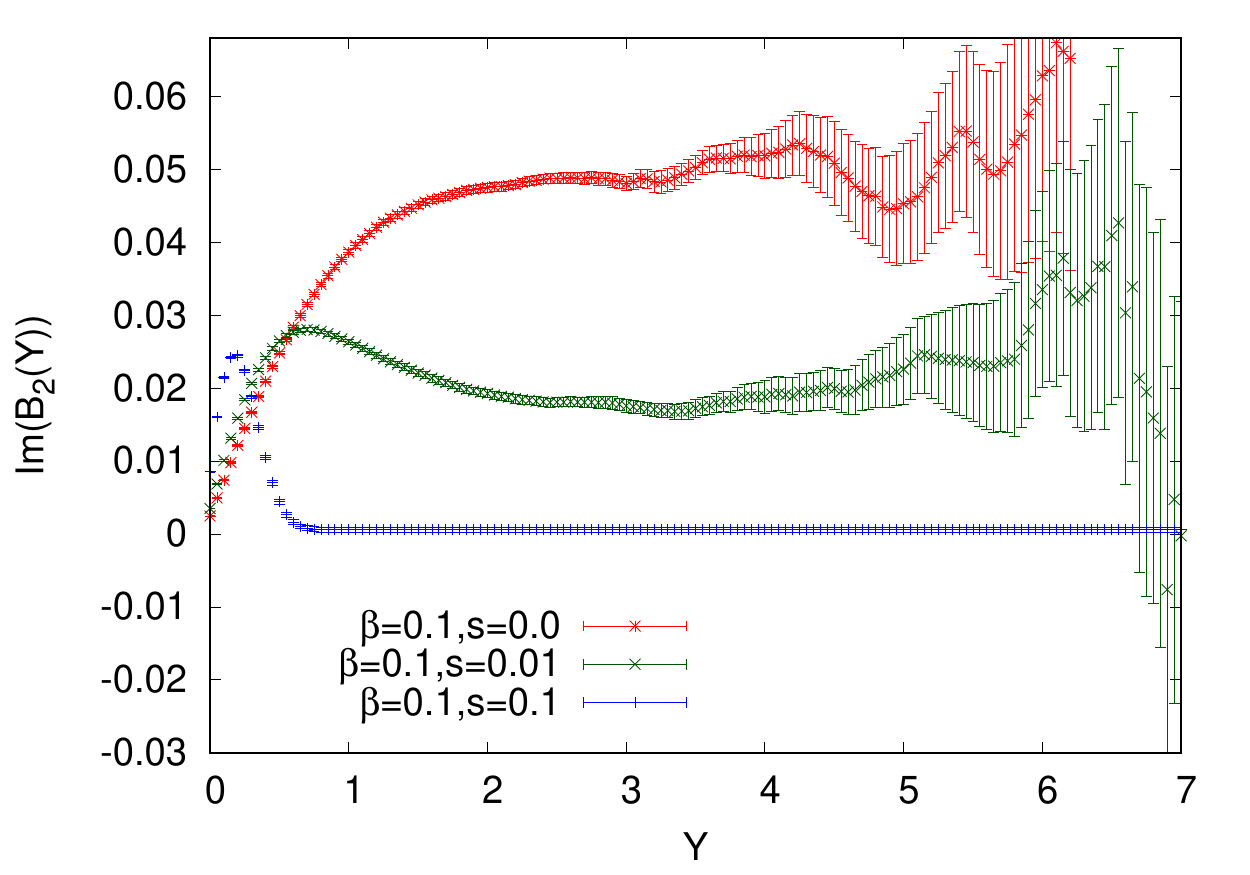}
%\vglue-20mm
  \caption{ The interpolation function $F(t,\tau)$ for the observable $e^{i\varphi} $
  in the toy model $S(\varphi) = i \beta \cos(\varphi) + s \varphi^2/2$, for
  $ \beta=0.1$ and various $t,\tau$ values (left).
  The second order boundary term $B_2$ for the observable $e^{i\varphi}$.
 }
\label{Ftt}
\end{center}
\end{figure}

First we investigate the toy model given by $ S(\varphi) = i \beta \cos(\varphi) + s \varphi^2 /2 $.
For $s=0$, the CLE equilibrates to $ P(x,y) = 1/(4 \pi \cosh^2 y ) $, giving
an incorrect value for the observables $ e^{i k \varphi} $ with
$ k \in \mathbb{Z} $ \cite{Salcedo:2016kyy}.
The second term in the action acts as a regularizer and ensures correct CLE results if $s$ is chosen
sufficiently large as can easily be verified using numerical integration.
The boundary term is measured using eq.~(\ref{Bncutoff}), as seen in Fig.~\ref{BT1}.
As one observes a nonzero $\lim_{Y\rightarrow \infty} B(Y)$ limit signals an incorrect CLE result.
Note that in the case of incorrect CLE results, the boundary term's fluctuations grow as
$Y$ grows. For the rightmost value on the plot the cutoff is removed, and the value $B(Y)$
is consistent with zero within errors, as argued above.
For this toy model, the whole $\tau$ dependence of $F(t,\tau)$ is calculable
using the solution of the Fokker-Planck equation for $P(x,y,t)$ and a numerical solution of
the differential eqs. defining $O(z,t)$ \cite{scherzer1}, see in Fig.~\ref{Ftt}.
As one observes, the $\tau$ dependence of $F(t,\tau)$ can be approximated with an ansatz.
\bea
 F(t,\tau) \approx  \sum_{n=0}^\infty A_n(t) \exp(-\omega_n \tau)
\eea
To get the error of the CLE solution, we must calculate
\bea
 F(t,0)- F(t,t) = \langle O \rangle_P - \langle O \rangle_\rho.
\eea
Using the simplified ansatz $F(t,\tau) = A_0 + A_1 \exp(-\omega_1 \tau) $, we can express
the error using the boundary terms:
\bea \label{correction}
F(t,0)-F(t,t) = \left. {  ( \partial_\tau F(t,\tau) )^2 \over \partial^2_\tau F(t,\tau) }
\right|_{\tau=0} = { B_1^2  \over B_2 }
\eea
This allows to calculate a corrected result with $ \langle O \rangle_\textrm{corr} =
\langle O \rangle_P - B_1^2/B_2  $.
On Fig.~\ref{Ftt} the second order boundary term is plotted. Compared to $B_1$ they have
larger fluctuations so large statistics is needed for a reliable measurement.
As shown in Table.~\ref{u1table} this allows the recovery of the correct result within
errors.

\begin{table}
\begin{tabular}{|l|c|c|c|c|c|}
  \hline
  $\beta, s$ &  $B_1$ & $B_2$ & CL & correct & corrected CL\\
  \hline
0.1,  0 & -0.04859(45)& 0.0493(11)&    -0.00115(45)&-0.05006 &-0.04901(62)\\
0.1,  0.01 & -0.01795(49)& 0.01801(80)&  -0.03318(50) & -0.05006 & -0.05106(40) \\
0.1,  0.1 & -0.00048(30)& 0.00057(35)& -0.04957(31)     &  -0.05006 &-0.04997(6)\\
\hline
0.5,  0 & -0.2474(11) & 0.237(11)& 0.00003(23)   & $-0.25815 $&  -0.258(11) \\
0.5,  0.3 & $-0.05309(86)$& 0.0552(51)& -0.19658(70)   &$ -0.23841$ &  $-0.2473(37)$ \\
  \hline
\end{tabular}
\caption{The estimation of the correct result
  using the correction formula (\ref{correction})
  for the toy model $S(\varphi) = i \beta \cos(\varphi) + s \varphi^2/2$  
  the imaginary part of the observable $e^{ix} $. }
  \label{u1table}
\end{table}

\begin{figure}[ht]
\begin{center}
  \includegraphics[width=0.48\columnwidth]{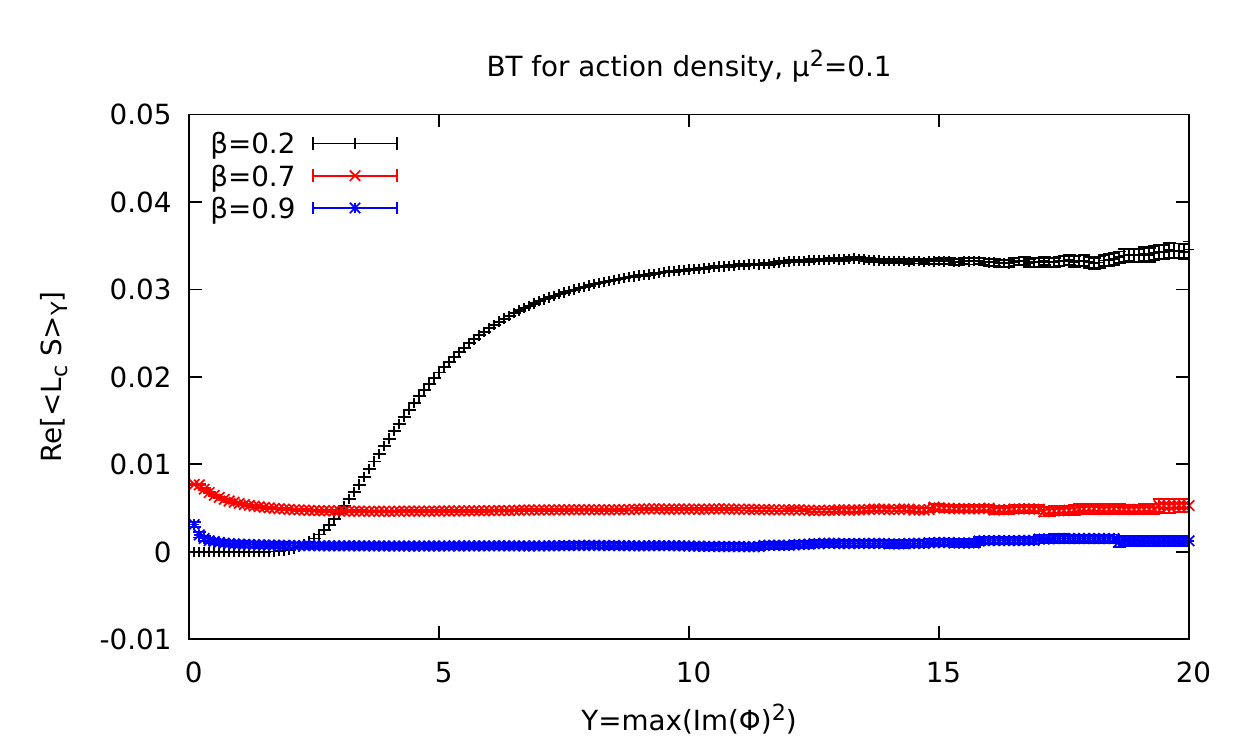}
  \includegraphics[width=0.48\columnwidth]{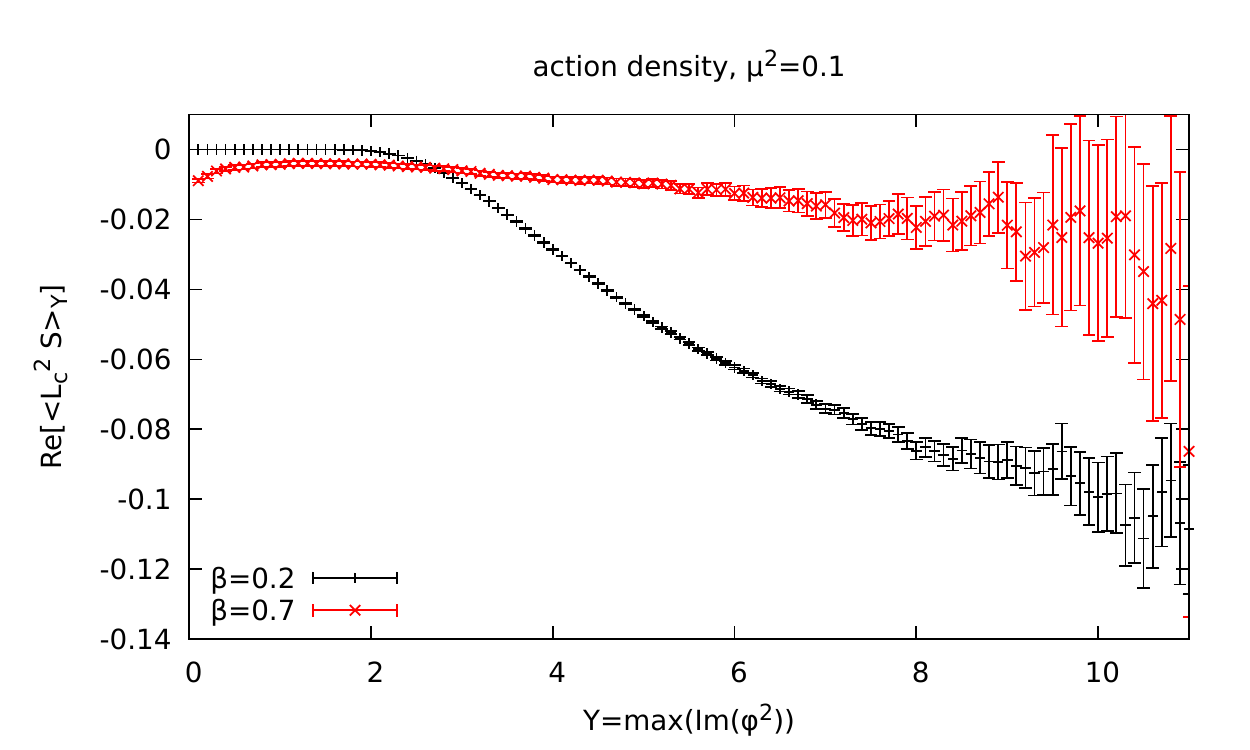}
%\vglue-20mm
  \caption{ The first order (left) and second order (right) boundary term in the 3D XY model for the action density observable.  }
\label{XYboundary}
\end{center}
\end{figure}

We have studied the boundary terms in the 3D XY model
\bea
 S = -\beta \sum_x \sum_{\nu=0}^2 \cos ( \phi_x - \phi_{x+\hat\nu} -i \mu \delta_{\nu,0} )
\eea
for which the CLE is known to fail in the small $\beta $ phase even for small
chemical potentials, and is apparently correct in the large $\beta$ region \cite{Aarts:2010aq}.
The measurement of the boundary terms confirms this picture, see in Fig.~\ref{XYboundary}.
Apparently a boundary term is present even at $\beta=0.9$, but its value is so small, that for
all intents and purposes the CLE gives correct results. The correction
of the observables using eq.~\ref{correction} gives the right magnitude and sign of the
systematic error of CLE, however the exact result (calculated using the worldline method)
in the low beta phase is not recovered. The reasons
for this might include the simplicity of the ansatz for $F(t,\tau)$ as well as the difficulty
of the measurement of the second order boundary terms due to lack of necessary statistics
(For further details, see \cite{scherzer2}).

\begin{figure}[ht]
\begin{center}
  \includegraphics[width=0.48\columnwidth]{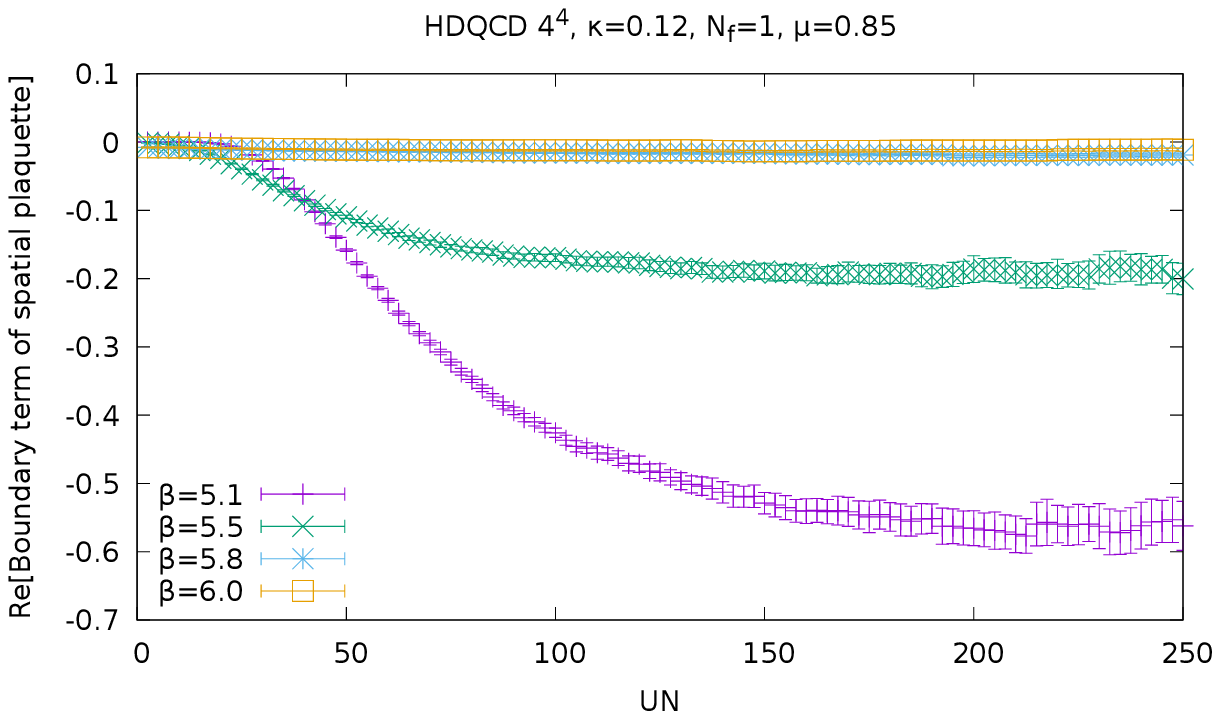}
  \includegraphics[width=0.48\columnwidth]{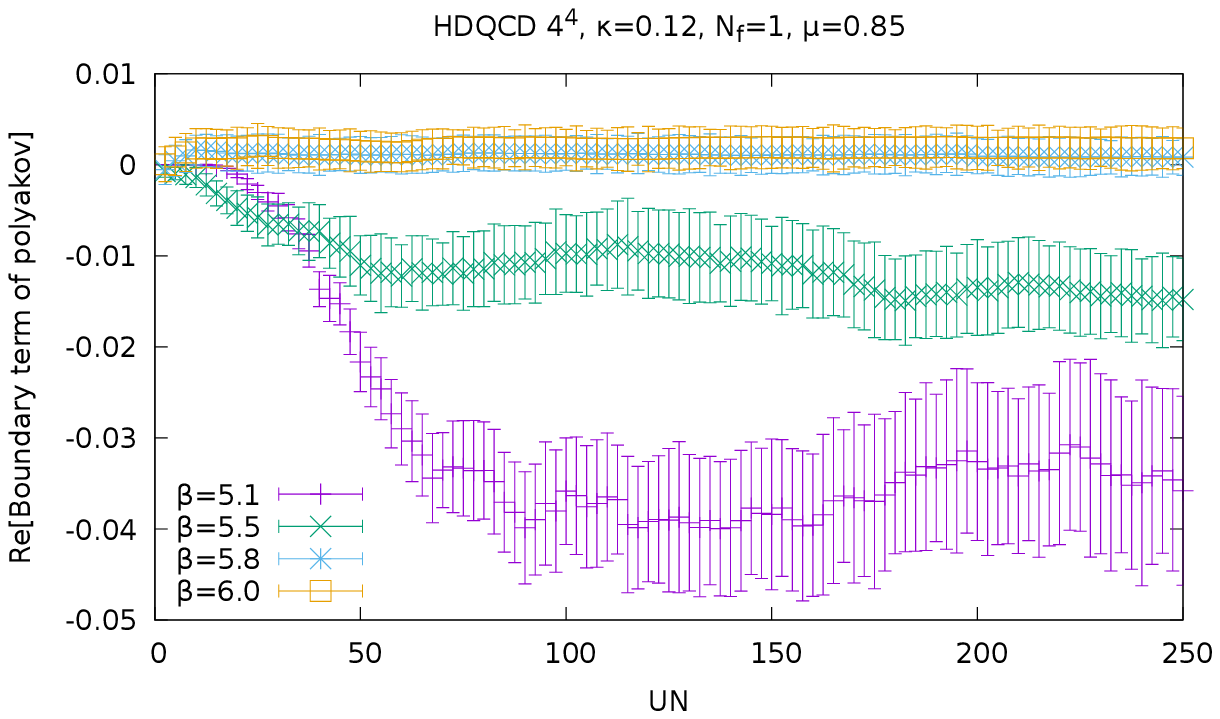}
%\vglue-20mm
  \caption{ The boundary term in HDQCD for the spatial plaquette average observable (left)
    and the polyakov loop observable (right).
    }
\label{HDQCDboundary}
\end{center}
\end{figure}
\begin{figure}[ht]
\begin{center}
  \includegraphics[width=0.48\columnwidth]{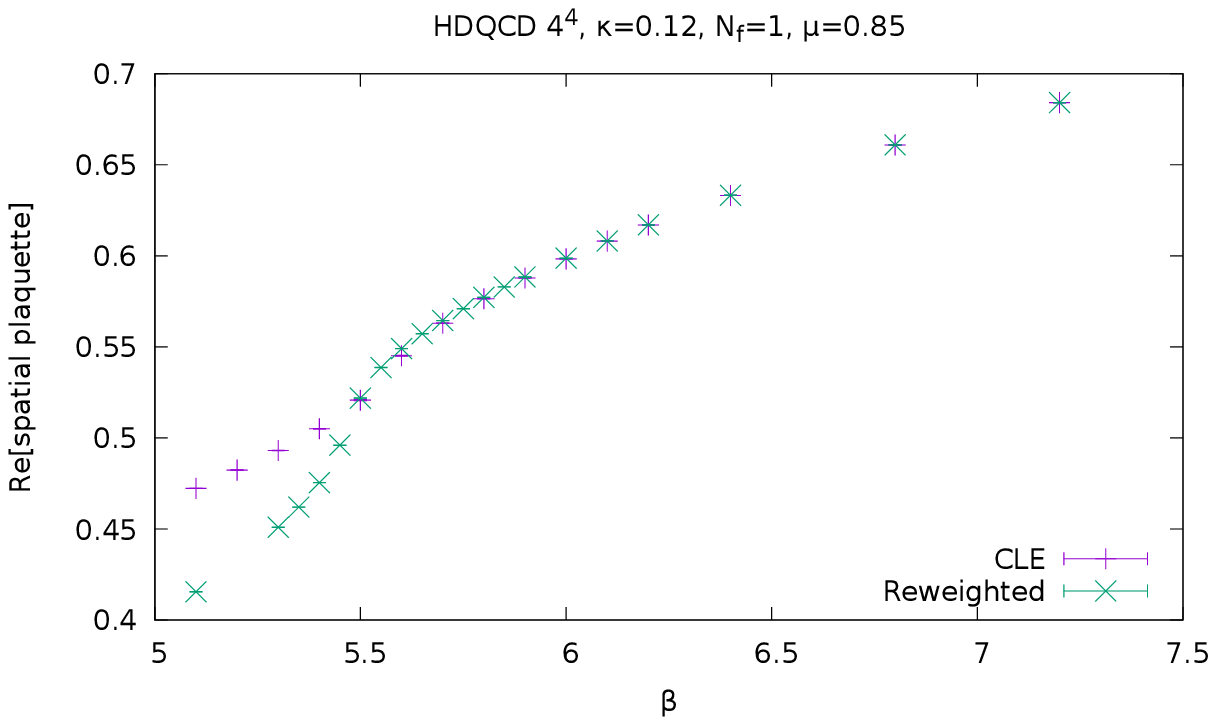}
  \includegraphics[width=0.48\columnwidth]{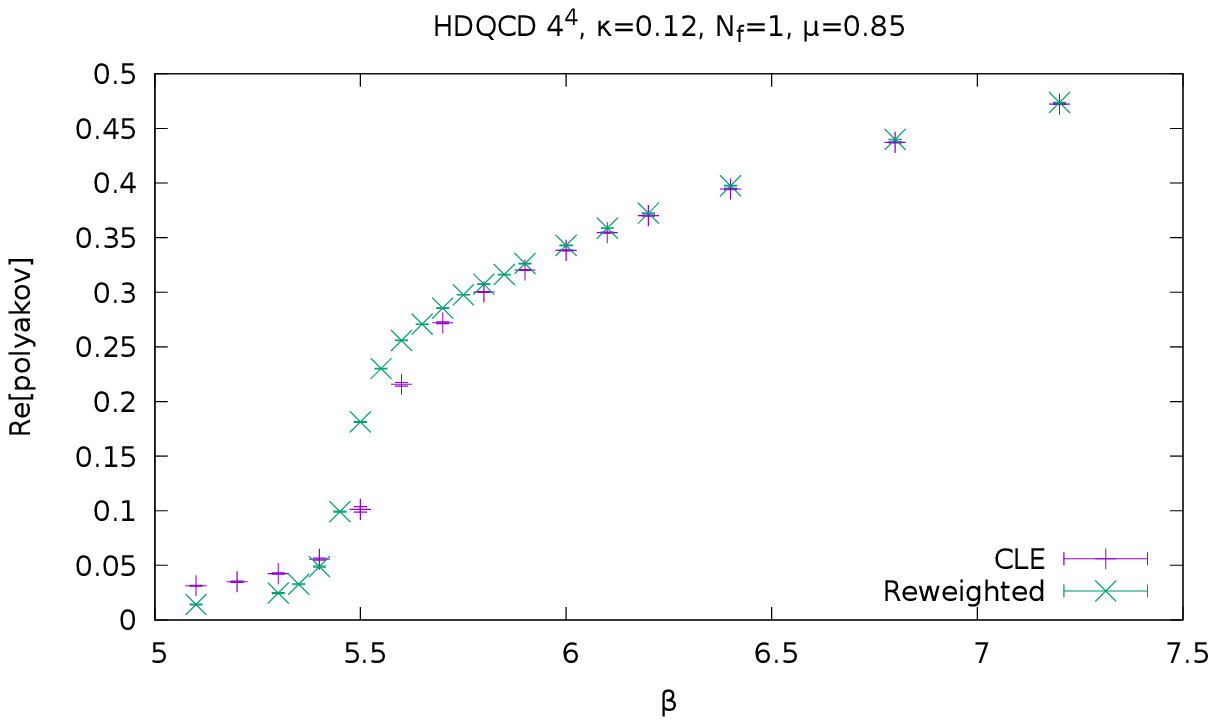}
%\vglue-20mm
  \caption{Comparing reweighting and CLE results in HDQCD: spatial plaquette (left) and Polyakov loop average (right). }
\label{HDQCDreweight}
\end{center}
\end{figure}

Finally we show results for the HDQCD model \cite{Bender:1992gn}. As noted in \cite{gaugecooling}
the CLE treatment gives correct results for large $\beta$ values, and incorrect result below
$ \beta \approx 5.8$.
The measurement of the boundary terms confirms this picture. In Fig.~\ref{HDQCDboundary}
the boundary term for the spatial plaquette and the Polyakov loop variable is shown.
In Fig.~\ref{HDQCDreweight} CLE results and reweighting results are shown, confirming
that it is possible to judge the reliability of CLE results based on the boundary terms.

\section{Conclusions}

We have shown that the boundary terms present a valuable diagnostic tool
for the assessing of the performance of a CLE simulation.
In the volume integral formulation, they can be measured
as the limiting value of a cut-off version of the observable $L_c O$ for the observable $O$.
The boundary term observables are cheap to measure even for lattice models, and one
needs a cheap offline analysis procedure to calculate the dependence on the cutoff.
The order of magnitude of the systematic error of CLE is than related to the magnitude
of the boundary terms. In some cases even the correction of the CLE result can be carried
out using second order boundary terms.

\end{document}